# Ground state with nonzero spontaneous magnetization of the two-dimensional spin-1/2 Heisenberg antiferromagnet with frustration

Toru Sakai,[1,2,a)] and Hiroki Nakano[1,b)]

[1]*Graduate School of Material Science, University of Hyogo, Kamigori, Hyogo 678-1297, Japan*

[2]*National Institutes for Quantum and Radiological Science and Technology (QST), SPring-8 Sayo, Hyogo 679-5148, Japan*

The $S = 1/2$ Heisenberg antiferromagnet on the two-dimensional pyramid lattice is studied by the numerical-diagonalization method. This lattice is obtained by the combination of the Lieb lattice and the square lattice. It is known that when interaction on the square lattice is increased from the ferrimagnetic limit of strong interaction on the Lieb lattice, this system shows gradual decrease and disappearance of spontaneous magnetization in the ground state. The present study treats the region near the case of the square-lattice antiferromagnet accompanied by isolated spins by numerical-diagonalization calculations of finite-size clusters with the maximum size of 39 sites. Our numerical results suggest the existence of a new phase with small but nonzero spontaneous magnetization between two zero-spontaneous-magnetization phases.

**I. INTRODUCTION**

Ferrimagnetism is a fundamental magnetic phenomenon which shows ferromagnetic nature and antiferromagnetic nature at the same time. It is widely known that the ferrimagnetism is mathematically understood by Marshall-Lieb-Mattis (MLM) theorem[1, 2]. The MLM theorem holds only when systems do not include magnetic frustration. It is a nontrivial problem how the ferrimagnetism is suppressed and disappears quantum-mechanically by such a frustrating situation. From such a viewpoint, there are several reports[3-17] which studied several models on lattices of various types. Some of them show intermediate states with spontaneous magnetization whose magnitude is smaller than that determined by the MLM theorem. Among systems showing spontaneously magnetized intermediate states, the triangular-lattice Heisenberg antiferromagnet with distortion of the $\sqrt{3}\times\sqrt{3}$ type shows a fascinating phenomenon[17-20]. The spontaneous magnetization decreases from the magnitude for the dice-lattice case[21] through smaller spontaneous magnetization to zero spontaneous magnetization for the undistorted triangular-lattice case. However, a recent report based on a numerical-diagonalization study[20] clarified the existence of another spontaneous-magnetization phase in the region of the interaction controlling the distortion is further varied. The new phase with spontaneous magnetization is different from the ferrimagnetic phase near the dice-lattice case and the previously reported spontaneous-magnetization phase neighboring the ferrimagnetic phase. One of the characteristic behaviors of the new phase is the discontinuity in the spontaneous magnetization at one boundary of the phase whereas the continuity at the other boundary. The nontrivial behavior of the distorted triangular-lattice system is confirmed by a density-matrix-renormalization-group study[22]. However, to understand the nontrivial behavior of the new spontaneous-magnetization state deeply, it is necessary to investigate another system showing the same behavior of such a spontaneous-magnetization phase.

---

[a]Electronic mail: sakai@spring8.or.jp.
[b]Electronic mail: hnakano@sci.u-hyogo.ac.jp

Under these circumstances, the purpose of this study is to present a new system that shows such a spontaneous-magnetization phase. The lattice of the system is composed of the Lieb-lattice and square-lattice networks of antiferromagnetic interactions[23]. This paper is organized as follows. In the next section, the target system is introduced. The method used in my calculations is also explained. The third section is devoted to the presentation and the discussion of numerical results. In the final section, a summary of this study is given.

## II. MODEL HAMILTONIAN AND METHOD

The Hamiltonian treated in this study is given by

$$\mathcal{H} = \sum_{i \in B, j \in B'} J_1 \, \mathbf{S}_i \cdot \mathbf{S}_j + \sum_{i \in A, j \in B} J_2 \, \mathbf{S}_i \cdot \mathbf{S}_j + \sum_{i \in A, j \in B'} J_2 \, \mathbf{S}_i \cdot \mathbf{S}_j \qquad (1)$$

where $S_i$ denotes the $S = 1/2$ spin operator at site $i$. The case of isotropic interaction in spin space is considered. The site $i$ is assumed to be the vertices of the Lieb lattice. The Lieb lattice is composed by interaction bonds $J_2$ when $J_1 = 0$. The vertices are divided into three sublattices A, B, and B'. Each site $i$ in the A sublattice is linked by four bonds of $J_2$; each site $i$ in the B or B' sublattice is linked by two bonds of $J_2$ and four bonds of $J_1$. The local structure in each unit cell of the lattice looks like a pyramid cone. The square bottom face of each cone shares its vertices with the pyramid cone in a neighboring unit cell. Each site of B and B' sublattice is a vertex of the two-dimensional square lattice composed of $J_1$ bonds. Therefore, we will call the present lattice the two-dimensional pyramid (2DP) lattice hereafter. Note also that the 2DP lattice seems similar to the checkerboard lattice[24-26]; the existence of A-site vertices in the present lattice is a significant difference from the checkerboard lattice. The number of spin sites is denoted by $N$; $N/3$ should be an integer because a unit cell includes three spins of A, B, and B' sublattices. We denote the ratio of $J_2/J_1$ by $r$. Energies are measured in units of $J_1$; hereafter, we set $J_1 = 1$. We consider that all interactions are antiferromagnetic, namely, $J_1 > 0$ and $J_2 > 0$. Note that for $J_2 \to \infty$, namely, $r \to \infty$, $J_1$ gets relatively smaller; the 2DP lattice structure then forms the so-called Lieb lattice. Numerical-diagonalization study[23] clarified that the decreasing $r$ makes the spontaneous magnetization get smaller gradually and disappear until $r \sim 2$. In the present study, on the other hand, we mainly investigate the region of $r \leq 2$.

The finite-size clusters that we treat in the present study are depicted in Fig. 1, namely, $N = 24, 27, 30,$ and 39. The periodic boundary condition is imposed for all $N$. Note here that all the clusters form a regular square although the clusters for $N = 24$, 30, and 39 are tilted. Since the cluster shape is a square, it is expected to capture the two-dimensionality of the system well. We calculate the lowest energy of the Hamiltonian in the subspace characterized by $M = \Sigma_i S^z_i$ by numerical diagonalizations based on the Lanczos algorithm and/or the Householder algorithm. The numerical-diagonalization method is unbiased against any approximations. Therefore, one can obtain reliable information of the system. The energy is denoted by $E(N, M)$, where $M$ takes an integer or a half odd integer up to the saturation value $M_{sat}(= NS)$. We define $M_{spo}$ as the largest value of M among the lowest-energy states, where $M_{spo}$ corresponds to spontaneous magnetization on which we focus our attention. Note that for cases of odd $N$, the smallest $M_{spo}$ cannot vanish; the result of $M_{spo} = 1/2$ in the ground state indicates that the system does not show a spontaneous magnetization. Part of the Lanczos diagonalizations were carried out using an MPI-parallelized code, which was originally developed in the study of Haldane gaps[27]. The usefulness of our program was confirmed in large-scale parallelized calculations[28-33].

## III. RESULTS AND DISCUSSIONS

First, let us observe our numerical results for the $M$-dependence of the lowest-energy level; the observation makes us understand how to determine the spontaneous magnetization for a given $N$. Figure 2 depicts results for $N = 30$ and 39. In Fig. 2(a) for $N = 30$, when $r = 1.3$, the energy for $M = 0$ is lower than any other energies for $M > 0$, which indicates that the spontaneous magnetization does not appear in the ground state, namely, $M_{spo} = 0$. For $r = 1.1$ and 0.9, on the other hand, the ground states are degenerate; the degeneracy is illustrated by broken lines in Fig. 2. These degeneracies indicate that nonzero $M_{spo}$ appears. The largest $M$ in the degenerated ground states corresponds to $M_{spo}$; one finds $M_{spo} = 1$ and $M_{spo} = 2$ for $r = 1.1$ and 1.3, respectively. In Fig. 2(b) for $N = 39$, on the other hand, possible values for $M$ are different; $M$ is supposed to be only a half odd integer. The levels for $M = \pm 1/2$ are necessarily degenerate even for arbitrary $r$; the degeneracy at $M = \pm 1/2$ for $r = 0.9$ corresponds to $M_{spo} = 1/2$ but does not indicate nonzero spontaneous magnetization as we mentioned above. For $r = 1.1$ and $r = 1.3$, the situation of the degeneracy is different from $r = 0.9$. Our numerical results indicate that $M_{spo} = 3/2$ and $M_{spo} = 5/2$ for $r = 1.1$ and 1.3, respectively.

Next, we examine the change of the spontaneous magnetization when $r = J_2/J_1$ is varied around the region below $r = 2$; results are depicted in Fig. 3(a). For $N = 24$, we do not find the region of nonzero spontaneous magnetization; results for $N = 24$. For larger $N$, on the other hand, we actually find the region of nonzero spontaneous magnetization. It is noticeable that two steps indicating nonzero spontaneous magnetization appear for $N = 30$ and 39 whereas only a step appears for $N = 27$. This behavior for larger systems suggests that the appearance of this region is not a finite-size effect.

Then, let us examine the boundary of the region of nonzero spontaneous magnetization. For this examination, with decreasing $r$, we define $r_{c2}$ as the value of r where $M_{\rm spo}$ changes from 0 or 1/2 to larger values and $r_{c1}$ as the value of $r$ where $M_{\rm spo}$ decreases again to 0 or 1/2 from larger values. The system size dependences of $r_{c1}$ and $r_{c2}$ are depicted in Fig. 3(b). One finds that both the dependences are small between $N = 30$ and 39 although the dependences between $N = 27$ and 30 are relatively larger. The small dependences for larger $N$ suggest that the width of this region survives when $N$ is increased. Therefore, these two panels of Fig. 3 gives a consequence that the region of nonzero spontaneous magnetization certainly exists as a macroscopic phenomenon in the thermodynamic limit.

Note here that the discontinuous behavior of $M_{\rm spo}$ at $r_{c1}$ appears whereas the continuous behavior at $r_{c2}$. To capture well the discontinuous behavior at $r_{c1}$, let us observe the $M$-dependence of the ground-state energy; our numerical results for $N = 30$ and 39 are depicted in Fig. 4. Figure 4(a) for $N = 30$ shows that for $r = 0.840$, one finds $M_{\rm spo} = 2$; however, our results for $r = 0.835$ and 0.830 indicate that $M_{\rm spo}$ vanishes. For $r = 0.835$, It is noticeable that the degeneracy from $M = 1$ and $M = 2$ still survives in an excited state. In our calculations for $r = 0.830$, on the other hand, the degeneracy disappears. Figure 4(b) for $N = 39$ shows that for $r = 0.85$, one finds $M_{\rm spo} = 5/2$; however, our results for $r = 0.84$ and smaller $r$ indicate that spontaneous magnetization disappears. Our results for $N = 39$ for $r = 0.81$ capture the behavior that the region of $M$ with degeneracy gets larger in the larger-side of M as well as the behavior of collapsing degeneracy becomes more marked in the smaller-side of $M$. One finds that the degeneracy gradually collapses from the side of smaller-$M$. The behavior around $r = r_{c1}$ is absolutely different from the behavior around $r = r_{c2}$ shown in Fig. 2.

**IV. SUMMARY**

We have studied the $S = 1/2$ Heisenberg antiferromagnet on the two-dimensional pyramid lattice by the numerical-diagonalization method. We find a region of nontrivial states showing spontaneous magnetization in the region near the case of the square-lattice system accompanied by isolated spins. When the strength of the frustrating interaction is increased from the square-lattice system accompanied by isolated spins, the spontaneous magnetization discontinuously appears at $r = r_{c1}$, but the spontaneous magnetization gradually decreases and continuously disappears at $r = r_{c2}$. The behavior of this system is similar to the one of the triangular-lattice system with distortion. Further comparison between these two cases contributes much for our understanding of frustration effects in magnetic materials.

**Acknowledgements**

We thank Professor N. Todoroki for the fruitful discussions. This work was partly supported by JSPS KAKENHI Grant Numbers 16K05418, 16K05419, 16H01080 (JPhysics), and 18H04330 (JPhysics). Nonhybrid thread-parallel calculations in numerical diagonalizations were based on TITPACK version 2 coded by H. Nishimori. In this research, we used the computational resources of the K computer provided by the RIKEN Advanced Institute for Computational Science through the HPCI System Research projects (Project ID: hp170018, hp170028, and hp170070). We also used the computational resources of Oakforest-PACS provided by JCAHPC through the HPCI System Research project (Project ID: hp170207 and hp180053). Some of the computations were performed using facilities of the Department of Simulation Science, National Institute for Fusion Science; Institute for Solid State Physics, The University of Tokyo; and Supercomputing Division, Information Technology Center, The University of Tokyo.

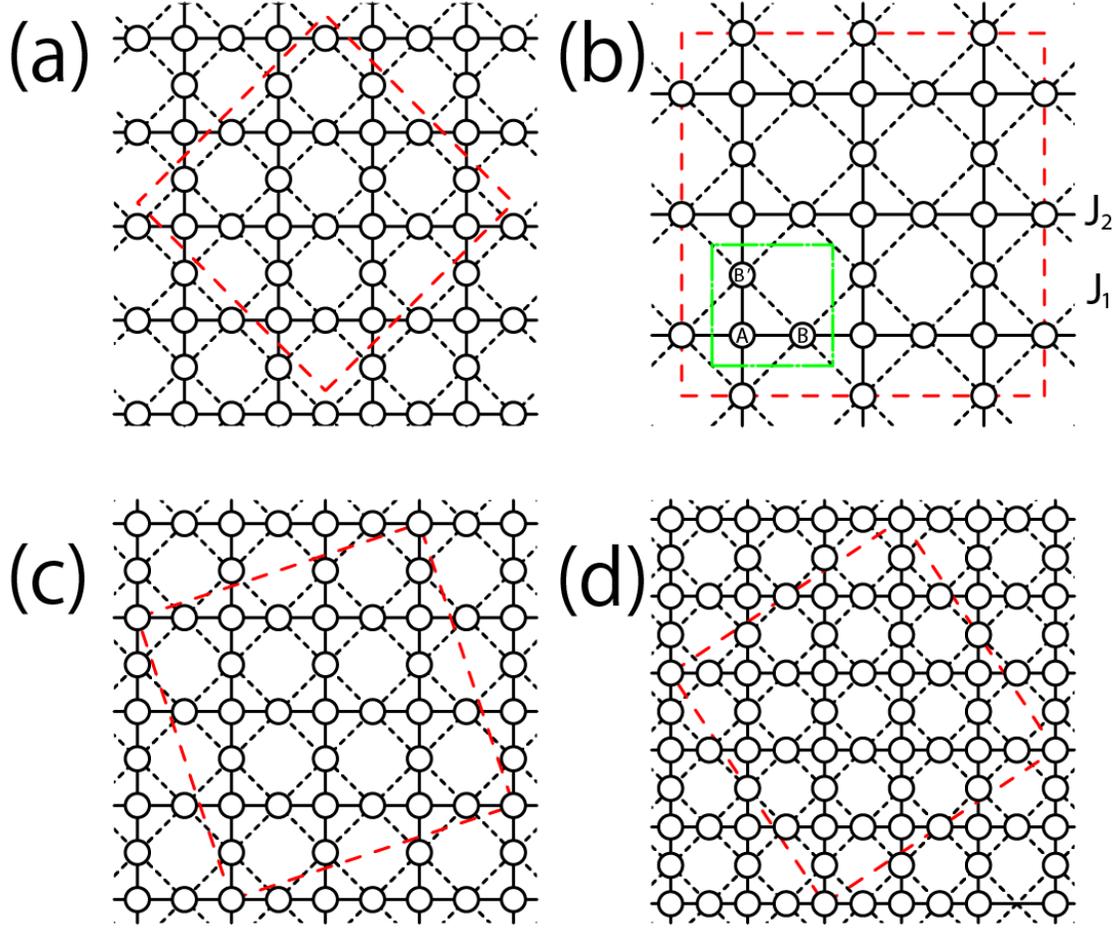

FIG. 1. Finite-size clusters treated in the present study. Panels (a), (b), (c), and (d) denote cases for $N = 24$, 27, 30, and 39, respectively by squares of red broken lines. In panel (b), a unit cell is represented by green dotted broken lines with sublattices. Bond strengths are also shown.

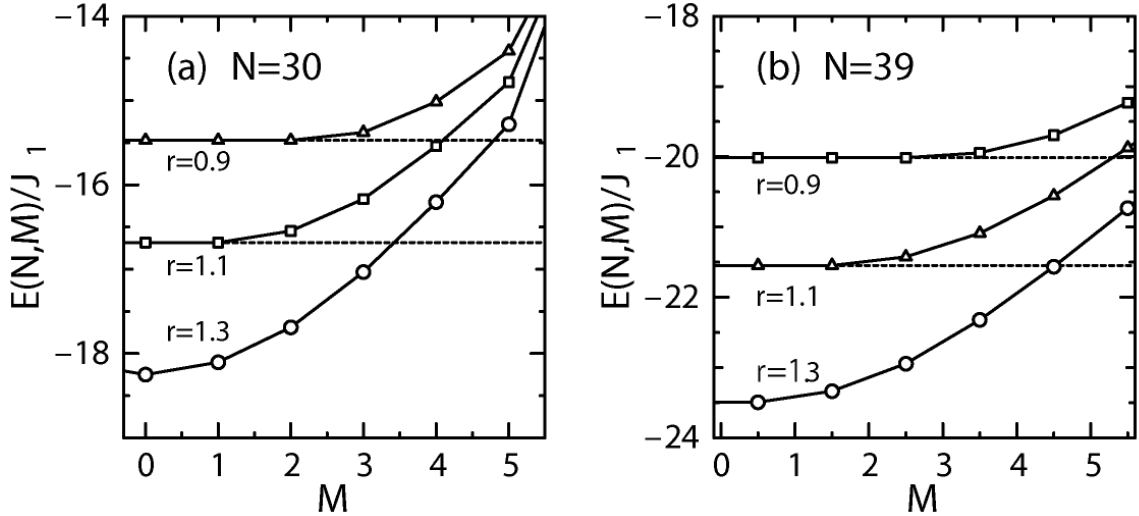

FIG. 2. $M$-dependence of the ground-state energy for $r = 0.9$, 1.1, and 1.3. Panels (a) and (b) denote cases for $N = 30$ and 39, respectively

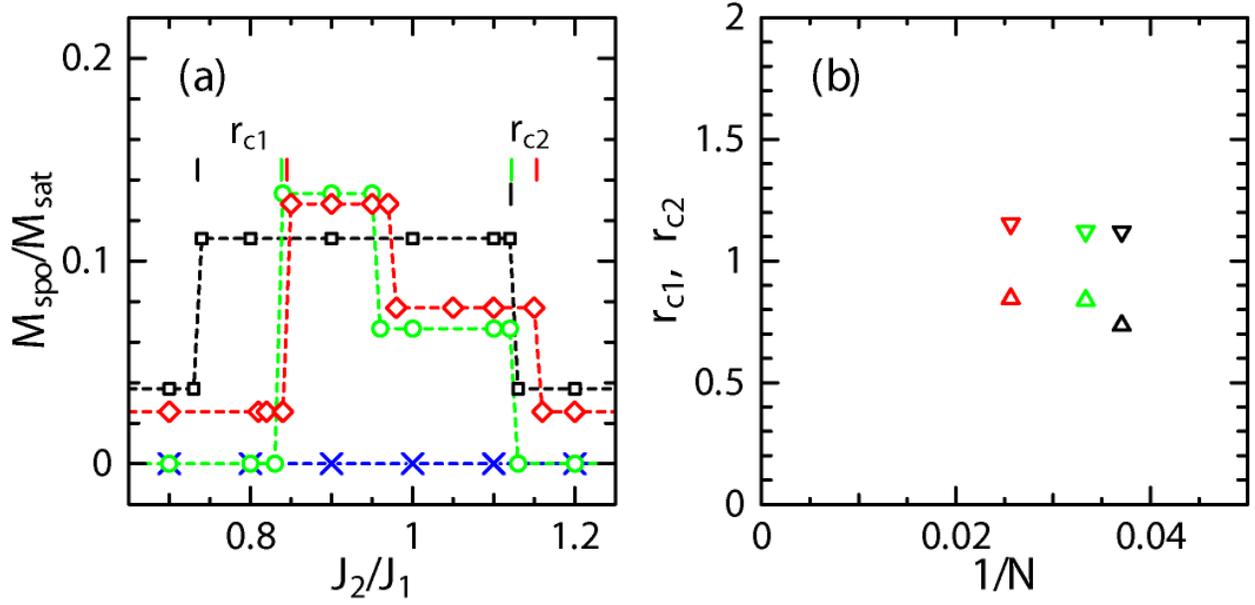

FIG. 3. (a) $r$-dependence of the spontaneous magnetization for various $N$. Blue crosses, black squares, green circles, and red diamonds denote results for $N = 24$, 27, 30, and 39, respectively. Results for $r_{c1}$ and $r_{c2}$ are shown with short solid lines with the corresponding color. (b) Size dependence of the boundaries of the nonzero-$M_{spo}$ phase. Triangles and inversed triangles denote results for $r_{c1}$ and $r_{c2}$, respectively.

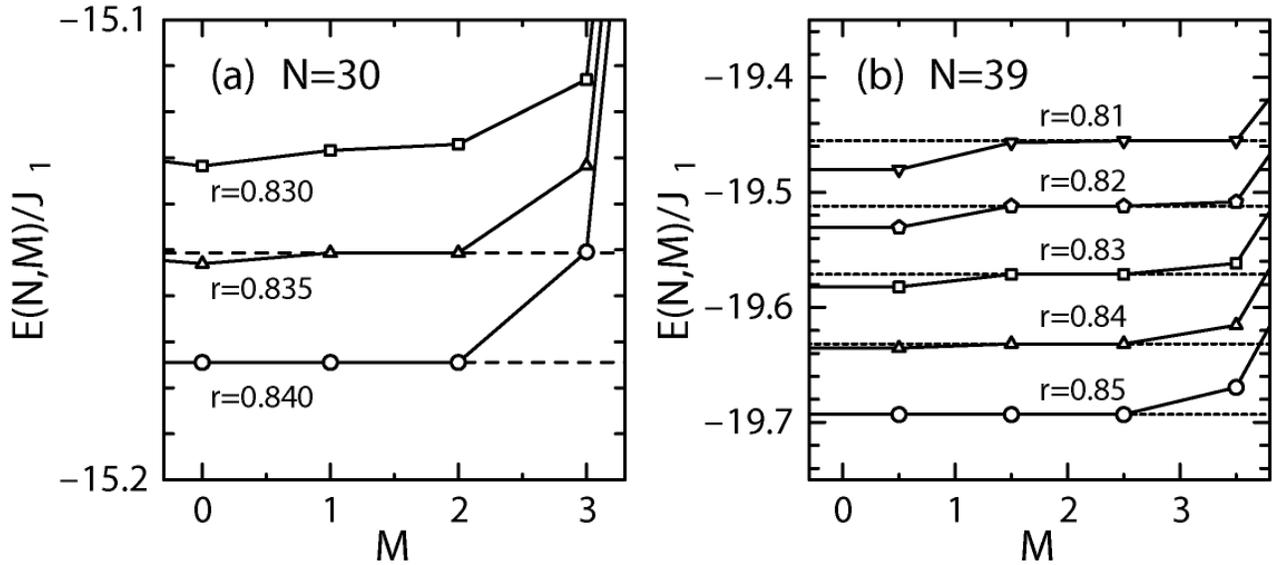

FIG. 4. $M$-dependence of the ground-state energy around $r_{c1}$. Panels (a) and (b) denote cases for $N = 30$ and 39, respectively.

8